\documentstyle[epsf,aps,prb]{revtex}

\begin{document}


\title{Magnetic and quantum disordered phases in triangular-lattice
Heisenberg antiferromagnets}

\author{ L. O. Manuel and H. A. Ceccatto}
\address{Instituto de F\'{\i}sica Rosario, Consejo Nacional de Investigaciones
Cient\'{\i}ficas y T\'ecnicas \\ and Universidad Nacional de Rosario, Bvd. 27
de Febrero 210 Bis, 2000 Rosario, Rep\'ublica Argentina }
\maketitle

\begin{abstract}

We study, within the Schwinger-boson approach, the ground-state structure of
two Heisenberg antiferromagnets on the triangular lattice: the $J_1%
\negthinspace-\negthinspace J_2$ model, which includes a
next-nearest-neighbor coupling $J_2$, and the spatially-anisotropic $J_1%
\negthinspace-\negthinspace J_1^{\prime }$ model, in which the
nearest-neighbor coupling takes a different value $J_1^{\prime }$ along one
of the bond directions. The motivations for the study of these systems range
from general theoretical questions concerning frustrated quantum spin models
to the concrete description of the insulating phase of some layered
molecular crystals. For both models, the inclusion of one-loop corrections
to saddle-point results leads to the prediction of nonmagnetic phases for
particular values of the parameters $J_1\negthinspace/\negthinspace J_2$ and
$J_1^{\prime }\negthinspace/\negthinspace J_1.$ In the case of the $J_1%
\negthinspace-\negthinspace J_2$ model we shed light on the existence of
such disordered quantum state, a question which is controversial in the
literature. For the $J_1\negthinspace-\negthinspace J_1^{\prime }$ model our
results for the ground-state energy, quantum renormalization of the pitch in
the spiral phase, and the location of the nonmagnetic phases, nicely agree
with series expansions predictions.
\end{abstract}



\section{Introduction}

The triangular-lattice Heisenberg antiferromagnet (TLHA) has played a
fundamental role in the understanding of frustrated quantum spin systems. In
particular, the existence of a nonmagnetic ground state in the $S=1/2$
system has been strongly debated in the literature, although there is now a
widespread conviction that it displays the classical 120$^{\circ }$ spiral
order.\cite{orden,Colo1} In the last years, this particular problem was
linked to more general questions concerning spin-liquid states and their
possible connections to high-$T_c$ superconductivity. In this context, the
most studied model has been the square-lattice Heisenberg antiferromagnet
with first- and second-neighbor interactions, the so-called $J_1%
\negthinspace -\negthinspace J_2$ model.\cite{12} The introduction of
frustrating second-neighbor interactions leads, at intermediate values of
the couplings, to the existence of a disordered spin-liquid state which
intervenes between the N\'eel and ''striped '' (ferromagnetic in one
direction, antiferromagnetic in the other) magnetic orders. This situation
is by now fairly well established by a variety of numerical and analytical
methods.\cite{12,letter} On the other hand, an extension of the TLHA
including second-neighbor interactions, the $J_1\negthinspace -\negthinspace %
J_2$ TLHA, has also been 
considered.\cite{TLHA12}$^{-}$\cite{Leche1} This extension is a
natural development after the thorough investigation of the model on the
square lattice, but the existence or not of an intermediate spin-liquid
phase in this case is not so clear, with most works favoring the
non-existence of such a state.\cite{Chu}$^{-}$\cite{Leche1} This is 
somehow paradoxical,
since the TLHA was considered for many years the best candidate to have a
spin-liquid ground state just on the basis of the frustrating lattice
topology. One would expect that the introduction of additional frustration
through the $J_{2\text{ }}$interaction should contribute to melt the already
weak order of the nearest-neighbor model.

The study of the $J_1\negthinspace -\negthinspace J_2$ TLHA was mostly
driven by purely theoretical motivations. On the other hand, recent
experimental results have produced a surge of 
interest\cite{McK}$^{-}$\cite{series} in
the ground-state properties of the nearest-neighbor TLHA with spatially
anisotropic couplings: $J_1$ along two bond directions and $J_1^{\prime }$
in the third one. It has been argued\cite{McK} that this model should
describe the magnetic phases of the quasi-two-dimensional organic
superconductors ${\it \kappa }-$(BEDT-TTF)$_2$X. For these layered molecular
crystals the relevant values of $J_1^{\prime }/J_1$ are around $0.3-1.0$,
with the ratio varying with the anion X and with uniaxial stress along the
layer diagonal.

We have previously performed a study of both the $J_1\negthinspace -%
\negthinspace J_2$ and $J_1\negthinspace -\negthinspace J_1^{\prime }$ TLHA
using the rotational invariant Schwinger-boson approach (SBA) in a
mean-field approximation.\cite{GC} At that time, our motivations for
studying the latter model was the natural interpolation that it provides
between the nearest-neighbor TLHA ($J_1^{\prime }=J_1$) and the
square-lattice antiferromagnet ($J_1^{\prime }=0$). We obtained a good
agreement between the ground-state energy predicted in our approach and
exact numerical values on finite lattices. Furthermore, no indication of a
disordered state was found for the values of $J_2/J_1$ and $J_1^{\prime
}/J_1 $ of interest. In this work we extend the calculations in \cite{GC} to
include one-loop corrections to the mean-field picture. We have shown\cite
{letter,fluctri} that these corrections bring the saddle-point results in
line with exact diagonalization values on finite clusters, which lends
support to the SBA predictions for the thermodynamic limit. In particular,
for the square-lattice $J_1\negthinspace -\negthinspace J_2$ model we found
that there is a quantum nonmagnetic phase for $0.53\negthinspace \lesssim 
\negthinspace J_2/J_1\negthinspace \lesssim \negthinspace 0.64$. This result
was obtained by considering the spin stiffness on large lattices and
extrapolating to the thermodynamic limit, a procedure that avoids the
infinite-lattice infrared divergencies associated to Bose condensation. We
show here that similar behaviors are predicted for both the $J_1%
\negthinspace -\negthinspace J_2$ and $J_1\negthinspace -\negthinspace %
J_1^{\prime }$ TLHA.

\section{The calculational method}

We consider a general Hamiltonian of the form 
\begin{equation}
\label{H}H=\frac 12\sum_{{\bf r,r^{\prime }}}J({\bf r-r^{\prime }})\ {\vec S}%
_{{\bf r}}\cdot {\vec S}_{{\bf r^{\prime }}}, 
\end{equation}
where ${\bf r,r^{\prime }}$\ indicate sites on a triangular lattice. As
usual, we write spin operators in terms of Schwinger bosons: \cite{A} ${\vec
S_i}\negthinspace =\negthinspace {\frac 12}{\bf a}_i^{\dagger }.{\bf \vec
\sigma }.{\bf a}_i$, where ${\bf a_i^{\dagger }}\negthinspace
=\negthinspace (a_{i\uparrow }^{\dagger },a_{i\downarrow }^{\dagger })$ is a
bosonic spinor, ${\bf \vec \sigma }$ is the vector of Pauli matrices, and
there is a boson-number restriction $\sum_\sigma a_{i\sigma }^{\dagger
}a_{i\sigma }\negthinspace =\negthinspace 2S$ on each site. With this
representation the spin-spin interaction can be written as ${\vec S_i}.{\vec
S_j}\negthinspace =:\negthinspace
B_{ij}^{\dagger }B_{ij}\negthinspace :-A_{ij}^{\dagger }A_{ij},$ where we
defined the $SU(2)-$invariant order parameters $A_{ij}=\frac 12\sum_\sigma
\sigma a_{i\sigma }a_{j{\bar \sigma }}$ and $B_{ij}^{\dagger }=\frac
12\sum_\sigma a_{i\sigma }^{\dagger }a_{j\sigma }\ ({\bar \sigma }=-\sigma
,\ \sigma =\pm 1)$, and the notation $:\negthinspace O\negthinspace :$
indicates the normal order of operator $O$. Thus, we can formulate the
partition function for the Hamiltonian (\ref{H}) as a functional integral
over boson coherent states, which allows its evaluation by a saddle-point
expansion. Since the theory presents a local $U(1)$ symmetry, we use
collective coordinate methods ---as developed in the context of relativistic
lattice gauge theories\cite{P}--- to handle the infinitely-many zero modes
associated to the symmetry breaking in the saddle point. These modes without
restoring forces, which correspond to local gauge transformations, are
eliminated by the exact integration of the collective coordinates. Such a
program can be carried out by enforcing in the functional integral measure
the so-called background gauge condition, or ``natural'' gauge, \cite{P} by
means of the Fadeev-Popoff trick. In this way we restrict the integrations
in the partition function to field fluctuations that are orthogonal to the
collective coordinates. At $T=0$, after carrying out the integrations on
these genuine fluctuations, we obtain the one-loop correction to the
ground-state energy {\it per site}, 
\begin{equation}
\label{E1}E_1=-\frac 1{2\pi }\int_{-\infty }^\infty d\omega \sum_{{\bf k}%
}\ln \left( \frac{\Delta _{{\rm FP}}({\bf k},\omega )}{|\omega |\sqrt{\det 
{\cal A}_{\perp }^{(2)}({\bf k},\omega )}}\right) . 
\end{equation}
Here the Fadeev-Popoff determinant $\Delta _{{\rm FP}}({{\bf k},\omega })%
\negthinspace =\negthinspace \left| {\vec \varphi }_0^L({{\bf k},\omega }).{%
\vec \varphi }_0^R({{\bf k},\omega })\right| ,$where ${\vec \varphi }_0^L({%
{\bf k},\omega })$ is the {\it left} zero mode of the dynamical matrix $%
{\cal A}^{(2)}$ in the ${\bf k}\negthinspace -\negthinspace \omega $
subspace, and ${\cal A}_{\perp }^{(2)}$ is the projection of this matrix in
the subspace orthogonal to the collective coordinates. The dynamical matrix $%
{\cal A}^{(2)}$ is the Hessian of the effective action as a function of the
decoupling (Hubbard-Stratonovich) fields (see \cite{letter} for details).

In the ordered phases of the model the Bose condensate breaks the global $%
SU(2)$ symmetry and its density gives the local magnetization.\cite{A} The
associated physical Goldstone modes at ${\bf k}\negthinspace =\negthinspace
0,\pm {\bf Q}$\ (${\bf Q}$ is the magnetic wavevector) lead to serious
infrared divergencies of intermediate quantities, which have to be cured by
standard renormalization prescriptions. In order to avoid these problems we
have computed physical quantities (which are free of divergencies) on large
by finite lattices, and finally extrapolated these values to the
thermodynamic limit. We considered clusters with the spatial symmetries of
the infinite triangular lattice, corresponding to translation vectors ${\bf T%
}_1=(n+m){\bf e}_1+m{\bf e}_2$, ${\bf T}_2=n{\bf e}_1+(n+m){\bf e}_2.$\cite
{Bernu,Leche2} 
Here ${\bf e}_1={\bf (}a,0{\bf )\ }$and ${\bf e}_2=(-%
{\bf \frac 12}a{\bf ,\frac{\sqrt{3}}2}a{\bf )\ }$are the basic triangular
lattice vectors.${\bf \ }$To fit to the cluster shapes the expected (spiral)
magnetic orders, and also to allow the calculation of the stiffness, we
impose arbitrary boundary conditions ${\vec S}_{{\bf r+T}_i}={\cal R}_{ 
\widehat{n}}(\Phi _i){\vec S}_{{\bf r}}$ ($i=1,2$) , where ${\cal R}_{ 
\widehat{n}}(\Phi _i)$ is the matrix that rotates an angle $\Phi _i$ around
some axis $\widehat{n}$ (notice that we are using boldface (arrows) for
vectors in real (spin) space). It is convenient to perform local rotations ${%
\vec S}_{{\bf r}}\rightarrow {\cal R}_{\widehat{n}}(\theta _{{\bf r}}){\vec S%
}_{{\bf r}}$ of angle $\theta _{{\bf r}}=\Delta {\bf Q\cdot r}${\bf ,} so
that with the choice $\Delta $${\bf Q\cdot T}_i=\Phi _i$ the boundary
conditions become the standard periodic ones ${\vec S}_{{\bf r+T}_i}={\vec S}%
_{{\bf r}}$. Then, we define the ($T=0$) stiffness tensor $\rho _{\widehat{n}%
}$ by 
\begin{equation}
\label{RHO}\rho _{\widehat{n}}^{ij}=\left. \frac{\partial ^2E_{{\rm GS}}(%
{\bf Q+}\Delta {\bf Q)}}{\partial \theta _i\partial \theta _j}\right|
_{\Delta {\bf Q=}0}, 
\end{equation}
where $E_{{\rm GS}}$ is the ground-state energy {\it per spin} and $\theta
_i=\Delta {\bf Q\cdot e}_i$ ($i=1,2$) are the twist angles along the basis
vectors ${\bf e}_i$. In the next two sections we will apply this formalism
to the models under consideration.

\section{The $J_1\negthinspace -\negthinspace J_2$ TLHA}

The $J_1\negthinspace -\negthinspace J_2$ TLHA is given by (\ref{H}) with $J(%
{\bf r-r^{\prime }})=J_1$($J_2$) for ${\bf r}$ and ${\bf r^{\prime }}$%
nearest (next-nearest) neighbor sites, and $0$ otherwise. For classical spin
vectors, when $\alpha \equiv J_2/J_1<1/8$ the lowest-energy configuration in
this model is the commensurate spiral with magnetic wavevector ${\bf Q}=( 
\frac{4\pi }{3a},0)$; for $1/8<\alpha <1$ there is a degeneracy between the
two-sublattice and four-sublattice N\'eel orders. Quantum fluctuations lift
this degeneracy and select, through an ''order from disorder '' phenomenon,
the two-sublattice collinear ground-state with magnetic wavevector ${\bf Q}%
=(\frac \pi a,-\frac \pi {\sqrt{3}a})$. This scenario was first proposed
using spin-wave theory\cite{Chu} and later confirmed by a study of the
thermodynamic-limit collapse to the ground state of low lying levels.\cite
{Leche1} The correction (\ref{E1}) to the saddle-point value $E_0$ leads to
the ground-state energy $E_{{\rm GS}}\negthinspace =\negthinspace E_0+E_1$
shown in Fig. 1. This figure contains the result for the infinite lattice
and also for a finite cluster of 12 sites, which allows a comparison with
exact results obtained by numerical diagonalization.\cite{SW} We see that
the addition of the Gaussian correction $E_1$ improves the saddle-point
value $E_0$, particularly for the 120$^{\circ }$ spiral phase. Moreover, the
departure of the fluctuation-corrected results from the exact values in the
range $0.1\lesssim\negthinspace \alpha \negthinspace \lesssim 0.2$ hints to
the possible existence of a disordered phase in this region. In the
thermodynamic limit, at saddle-point order the theory predicts a first order
transition between the two magnetic ground states at some intermediate
frustration $\alpha \simeq 0.16$, with no intervening disordered phase.\cite
{GC} After the inclusion of the Gaussian fluctuations there is a window $0.12%
\lesssim\negthinspace \alpha \negthinspace \lesssim 0.19$ where the
stiffness vanishes (see below) and the magnetic order is melted by the
combined action of quantum fluctuations and frustration.This result should
be compared with the linear spin wave results of  \cite{Iva}, which predict a
quantum nonmagnetic phase in the range $(0.1,0.14)$. Notice, however, that
within spin-wave theory this window closes when corrections to the
leading-order calculations are included.\cite{Chu} We stress that the same
happens in this approach for the $J_1\negthinspace -\negthinspace J_2$ model
on the square lattice, where however other methods confirm the existence of
a nonclassical phase.

As stated above, the existence or not of magnetic long-range order in the
thermodynamic limit was investigated by considering the spin-stiffness
tensor (\ref{RHO}). \cite{stiff} For spins lying on the $xy$-plane, it is
necessary to consider both the parallel stiffness $\rho _{\Vert }\equiv \rho
_{\widehat{z}}$ under a twist around $\widehat{z}$ and the perpendicular
stiffness $\rho _{\bot }\equiv \overline{\rho }_{xy}$ for twists around an
arbitrary versor $\widehat{n}$ on this plane. However, on clusters with
periodic boundary conditions our approach is rotational invariant and we
only have access to $\overline{\rho }=\frac 13(\rho _{\Vert }+2\rho _{\bot
}) $. On the other hand, we found that the large-$S$ Schwinger-boson
prediction for this quantity is exactly 4/3 smaller than the corresponding
classical result. As discussed in \cite{fluctri}, to solve these problems we
considered clusters that fit the magnetic orders in the $xy$-plane with
appropriate twisted boundary conditions.\cite{Bernu,Leche2} In this case,
since the rotational invariance is explicitly broken by the boundary
conditions, one is able to compute the parallel stiffness $\rho _{\Vert }$
and, moreover, the large-$S$ predictions have the correct behavior (no {\it %
ad hoc} factors are required to correct this quantity).\cite{fluctri}
Finally, using the values for $\overline{\rho }$ obtained from clusters with
periodic boundary conditions it is possible to determine $\rho _{\bot
}\equiv \frac{3.}2$ $\overline{\rho }-\frac 12\rho _{\Vert }$. The
corresponding results are presented in Fig. 2, both at saddle-point and
one-loop orders for comparison. Notice that in the 120$^{\circ }$ spiral
phase $\rho _{\Vert }$ softens first than $\rho _{\bot }$, while the
behavior is the opposite for the colinear state. Since the indirect
calculation of $\rho _{\bot }$ requires the use of {\it ad hoc} factors to
renormalize $\overline{\rho }$, the determination of the upper limit for the
disordered phase might be unreliable. However, based on our previous
experience with the SBA, we believe the existence of an intermediate
nonmagnetic phase can be trusted.

\section{The $J_1\negthinspace-\negthinspace J_1^{\prime }$ TLHA}

For the $J_1\negthinspace -\negthinspace J_1^{\prime }$ TLHA, in (\ref{H})
we take $J({\bf r-r^{\prime }})=J_1$ for ${\bf r-r^{\prime }=e}_{i\text{ }}$(%
$i=1,2$) and $J({\bf r-r^{\prime }})=J_1^{\prime }$ for ${\bf r-r^{\prime }=e%
}_3\equiv {\bf e}_1+{\bf e}_2.$ As mentioned above, this model is
interesting in view of its connections to the spin degrees of freedom in the
insulating phase of some layered organic superconductors.\cite{McK} Very
recent works have investigated its ground-state phase diagram and other
properties, using spin-wave theory\cite{Colo2} and series expansions.\cite
{series} For classical vectors, with $\eta \equiv J_1^{\prime }/J_1<1/2$ the
lowest-energy configuration is the two-sublattice N\'eel order discussed in
the previous section, with the (frustrated) ferromagnetic arrangement of the
spins along the weakly-coupled ${\bf e}_3$ direction. For $\eta >1/2$ the
preferred spin configuration becomes an incommensurate spiral, with the
angle $Q{\rm _{Class}}$ between neighboring spins along the ${\bf e}_1,{\bf e%
}_2$ directions given by $Q{\rm _{Class}}=\arccos (-\frac 1{2\eta }).$

The inclusion of the quantum nature of spins by means of the SBA produces
the results for the ground-state energy shown in Fig. 3. In this figure we
give the mean-field values and the fluctuation-corrected results, and
compare them with the series expansion predictions from \cite{series}. We
see that the fluctuations produce regions in which the saddle-point
solutions become unstable. On the other hand, for those values of $\eta $
where the magnetic phases considered are stable there is a very good
agreement between both methods. It is also of interest to compare the
quantum renormalization of the spiral pitch; in Fig. 4 we plot the classical
result for $Q$ given above, and the corresponding angle that minimize the
quantum ground-state energy. Again, in the region where the magnetic phases
are stable the results nicely agree with those coming from series
expansions. Furthermore, we have checked at mean-field order that the angle
that minimizes the total energy corresponds also to the minimum of the
quasiparticle dispersion relation, a fact that was used in \cite{series} to
determine $Q.$

To establish the regions without magnetic order we studied again the
spin-stiffness behavior. Like for classical vectors, the stiffness tensor (%
\ref{RHO}) is diagonal in the (perpendicular) directions ${\bf e}_3$ and $%
{\bf e}_{2}-{\bf e}_1$; the corresponding stiffness along these directions
are plotted in Fig. 5. We found a qualitatively different behavior between
the saddle-point and one-loop results; most notably, in the N\'eel phase the
Gaussian fluctuations soften first the classically stronger stiffness in the 
${\bf e}_3$ direction. At mean-field order our calculations indicate a
continuous transition between the two magnetic phases at $\eta \simeq 0.621$%
, and the absence of a magnetic saddle-point solution beyond $\eta \simeq
2.20$. The corrected results show the melting of the N\'eel phase at $\eta
\simeq 0.61$, still above the classical point $\eta {\rm _{Class}}=1/2$, and
predict a disordered quantum phase in the range $0.61\lesssim \eta \lesssim %
0.96$. Furthermore, the incommensurate phase becomes stable only in a
reduced parameter region $0.96\lesssim \eta \lesssim 1.10.$ In this case the
instability appears as a negative eigenvalue of the projected dynamical
matrix ${\cal A}_{\perp }^{(2)}$ in (\ref{E1}). These results are in fair
agreement with the series-expansion predictions from \cite{series}, which
indicate that the N\'eel order disappears at $\eta \sim 0.65-0.7$, the
disorder region extends from this value up to $\eta \sim 0.9$, and there is
an incommensurate phase for $\eta \gtrsim 0.9$ with no clear ending point.

\section{Conclusions}

In conclusion, we have considered, within the SBA, the Gaussian-fluctuation
corrections to the spin stiffness of the $J_1\negthinspace-\negthinspace J_2$
and $J_1\negthinspace-\negthinspace J_1^{\prime }$ TLHA. For the $J_1%
\negthinspace-\negthinspace J_2$ model we found that the order-parameter
fluctuations weaken the stiffness, which is reduced by increasing the
frustration until it vanishes leaving a small window $0.12\lesssim \alpha 
\lesssim 0.19$ where the system has no long-range magnetic order. Like in
previous studies, we found that the consideration of clusters which require
twisted boundary conditions to fit the magnetic orders avoids the use of 
{\it ad hoc} factors to correct the Schwinger-boson predictions. This fact
points to a subtle interplay between rotational invariance and the
relaxation of local constraints in this approach.

In the case of the $J_1\negthinspace-\negthinspace J_1^{\prime }$ TLHA our
results indicate a rich phase diagram, with at least two magnetic (N\'eel
and incommensurate spiral) phases and two disordered quantum states in the
parameter region of interest. One of the most notable aspects of our
calculations is the quantum renormalization of the magnetic wavevector in
the spiral phase, which agrees remarkably with the series expansion
prediction. We also found that the N\'eel order extends beyond its classical
stability point up to a value $\eta \simeq 0.61,$ where it seems to melt
continuously into a purely quantum phase. On the contrary, the spiral order
is favored in a reduced parameter range, with our results indicating first
order transitions from this phase to the nonmagnetic states. These first
order transitions appear in our calculations as local instabilities against
the order-parameter fluctuations.

Finally, there still remains the difficult task of characterizing the
disordered quantum phases of the models under consideration. Some 
attempts in this
direction have already been done, for both the $J_1\negthinspace -%
\negthinspace J_2$ \cite{TLHA12} and $J_1\negthinspace -\negthinspace %
J_1^{\prime }$ \cite{series} TLHA, but there is not clear understanding of
these phases yet. They are usually considered to be of the
resonant-valence-bond type, and are in general described starting from a
regular strong-bond (dimer) covering of the lattice.\cite{series} These
investigations can be performed within the formalism developed here, since
the SBA does not rely on having magnetic order in the system like, for
instance, spin-wave theory. However, these studies would require the
extension of the present calculations to larger magnetic unit cells and the
computation of physical quantities able to characterize the new phases, a
question that is far from trivial.

\acknowledgements
{We are grateful to Adolfo E. Trumper for useful discussions and for 
calling our attention to Ref. 12.}

\vskip 2cm

\figure{FIG. 1: Ground-state energy {\it per site} $E_{\rm GS}$ of 
the $J_1 - J_2$ TLHA as a function of the frustration parameter $\alpha=J_2/J_1$.
Dashed and full lines give the mean-field and fluctuation-corrected results, respectively.
The lower curves correspond to the energy of a 12-site cluster (the values 
are shifted by -0.05 for clarity); the upper curves give the thermodynamic-limit  results.
Dots are exact numerical values from \cite{SW}.}

\figure{FIG. 2: a) Parallel stiffness $\rho_{\Vert}$, and b) perpendicular 
stiffnes $\rho_{\perp}$ for the $J_1-J_2$ TLHA as  a function of the 
frustration parameter $\alpha$. Dashed 
and full lines give the saddle-point and fluctuation-corrected results, respectively. The thin 
vertical lines separate the regions with magnetic order from the middle parameter range 
where there is a quantum disordered phase.}

\figure{FIG. 3: Ground-state energy {\it per site} $E_{\rm GS}$ of 
the $J_1 - J^{'}_1$ TLHA as a function of $\eta=J^{'}_1/J_1$.
Dashed and full lines give the mean-field (MF) and fluctuation-corrected (FL)
results, respectively. Dots are series expansion predictions from \cite{series}.}

\figure{FIG. 4: Relative angle $Q$ between nearest-neighbor spins as a function of $\eta$. 
The thick full line corresponds to the lowest-energy configuration of classical spins, the 
dashed line is the mean-field prediction in the quantum case, and the thin full line is the 
series expansion result from \cite{series}. Open dots are the fluctuation-corrected 
results in the incommensurate phase (shown in more detail in the inset).}

\figure{FIG. 5: Spin stiffness $\rho_{\Vert}$ for the 
$J_1\negthinspace-\negthinspace J_1^{'}$ model as a function of $\eta$.
Dashed and full lines correspond
to saddle-point and fluctuation-corrected results, respectively. The thin vertical lines 
separate the regions with antiferromagnetic (AF) and incommensurate (INC) spiral
orders from the parameter ranges corresponding to disordered quantum phases.}

\newpage

\begin{figure}[ht]
\begin{center}
\epsfysize=10cm
\leavevmode
\epsffile{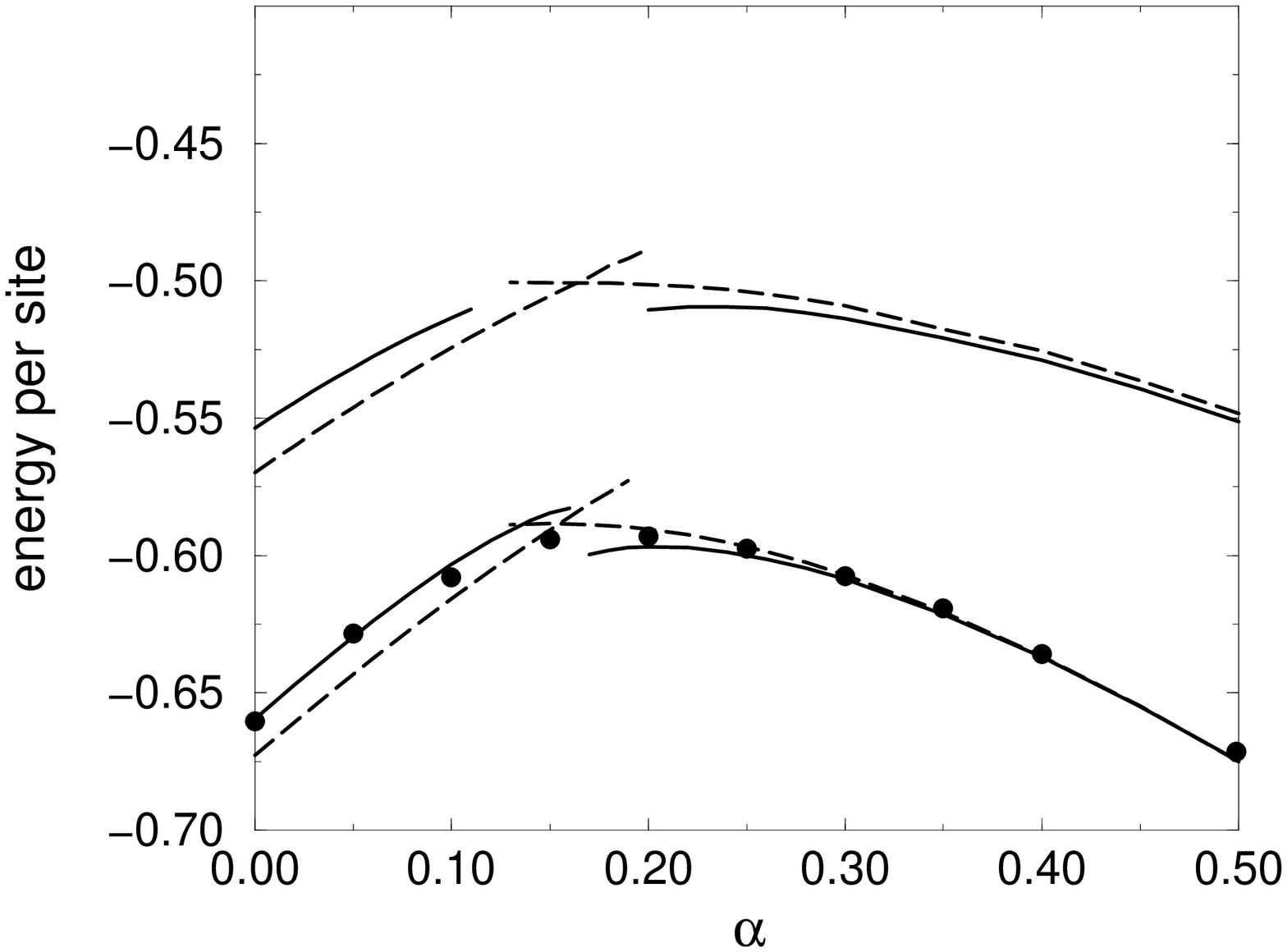}
\caption{}
\end{center}
\end{figure}

\begin{figure}[ht]
\begin{center}
\epsfysize=10cm
\leavevmode
\epsffile{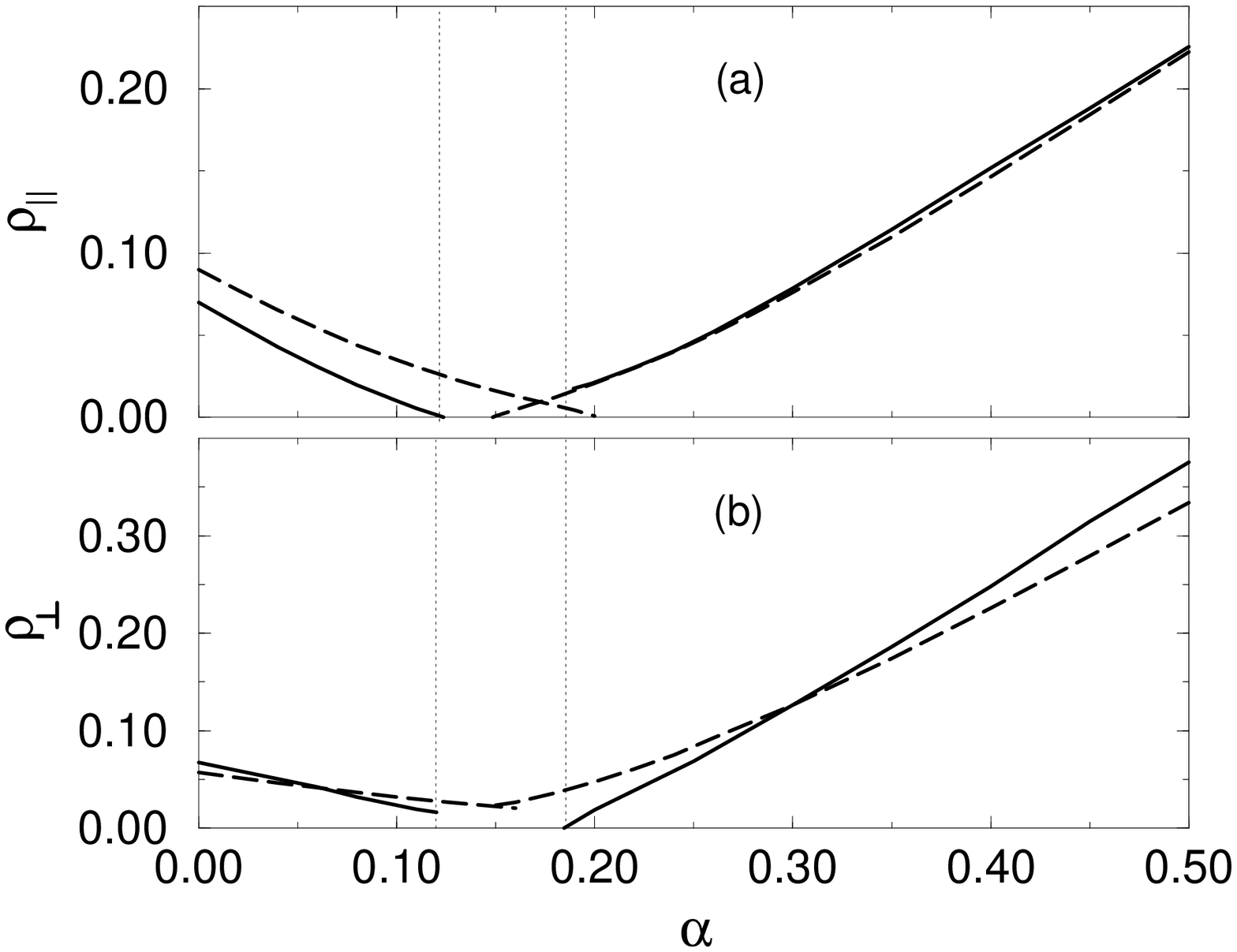}
\caption{}
\end{center}
\end{figure}

\begin{figure}[ht]
\begin{center}
\epsfysize=10cm
\leavevmode
\epsffile{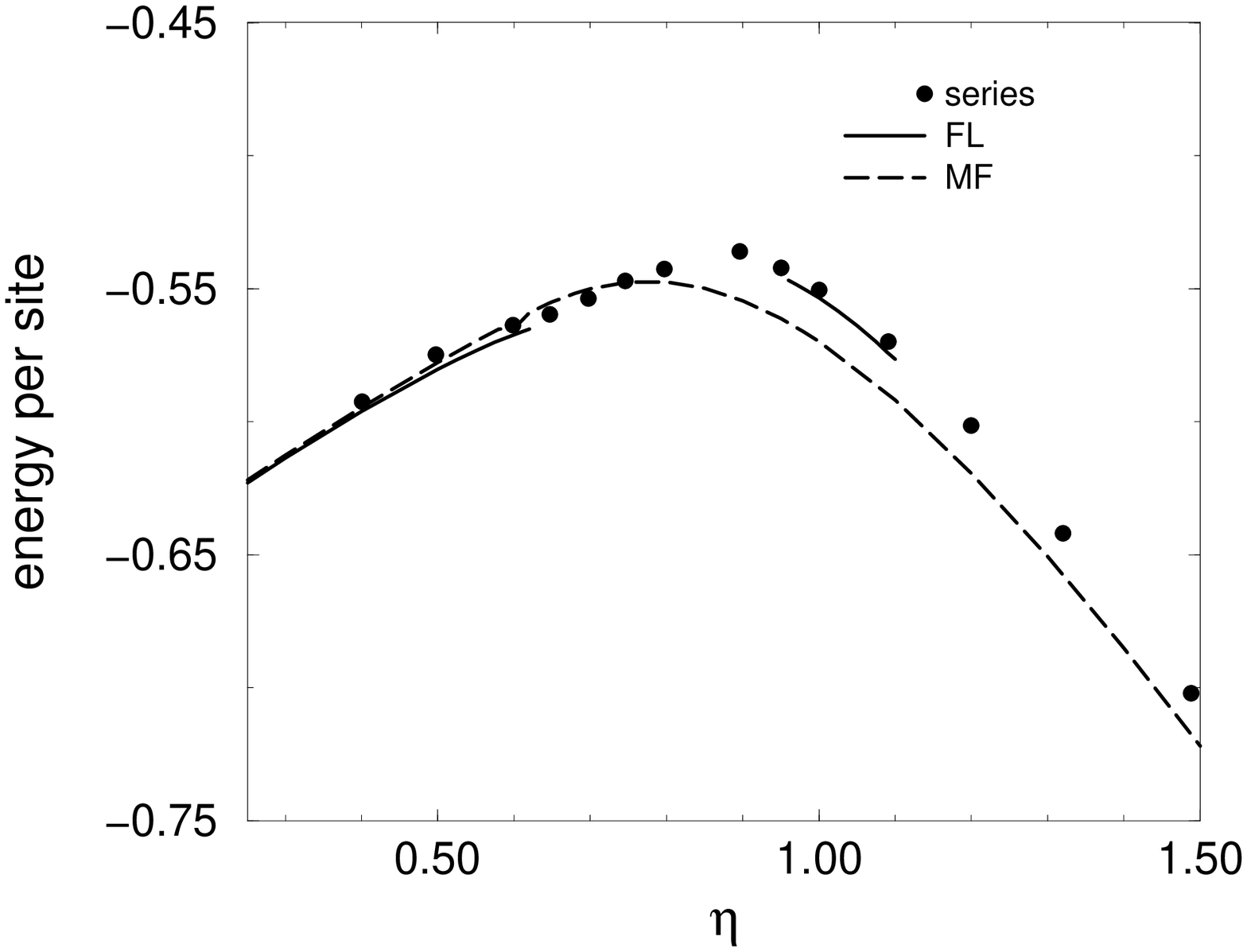}
\caption{}
\end{center}
\end{figure}

\begin{figure}[ht]
\begin{center}
\epsfysize=10cm
\leavevmode
\epsffile{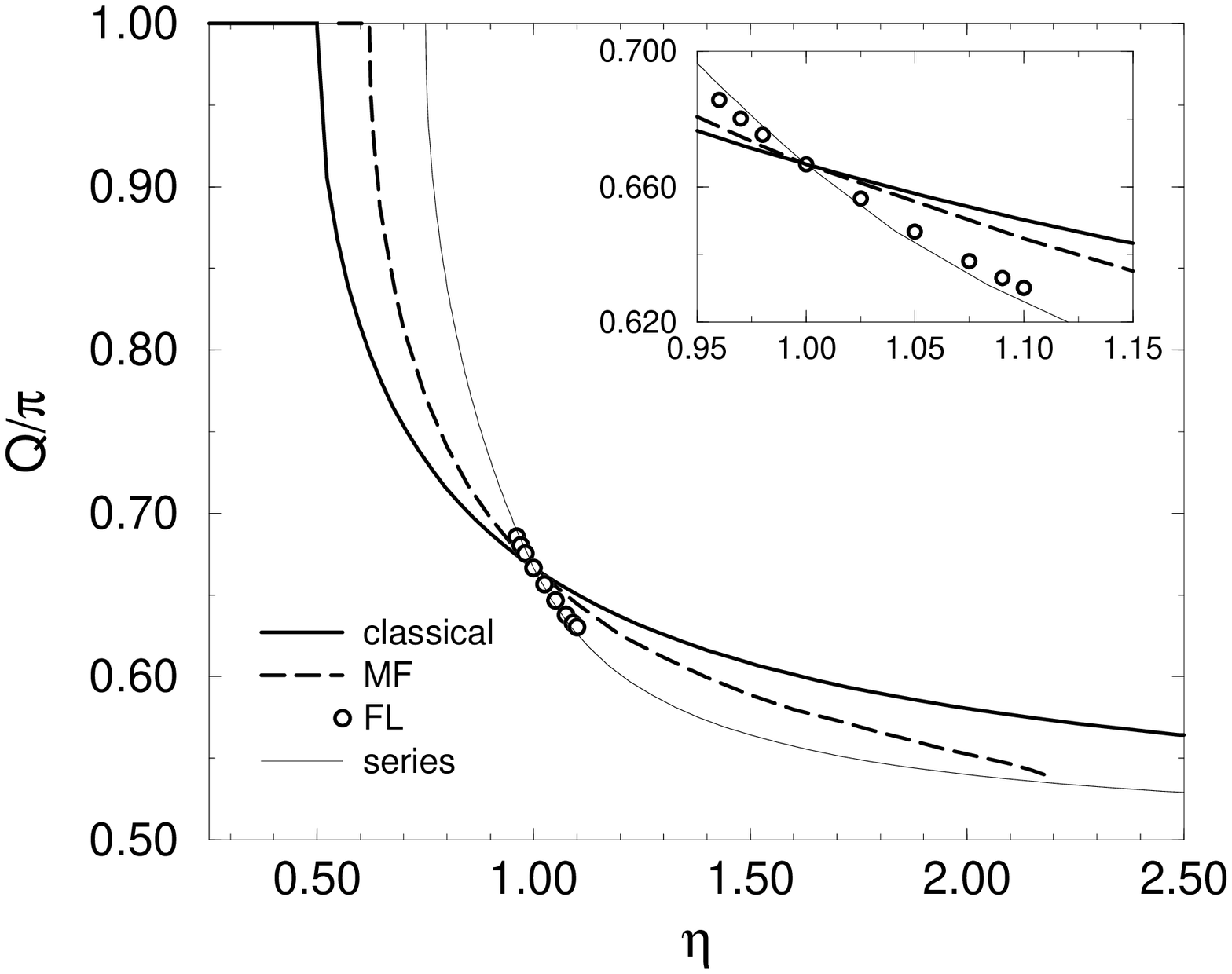}
\caption{}
\end{center}
\end{figure}

\begin{figure}[ht]
\begin{center}
\epsfysize=10cm
\leavevmode
\epsffile{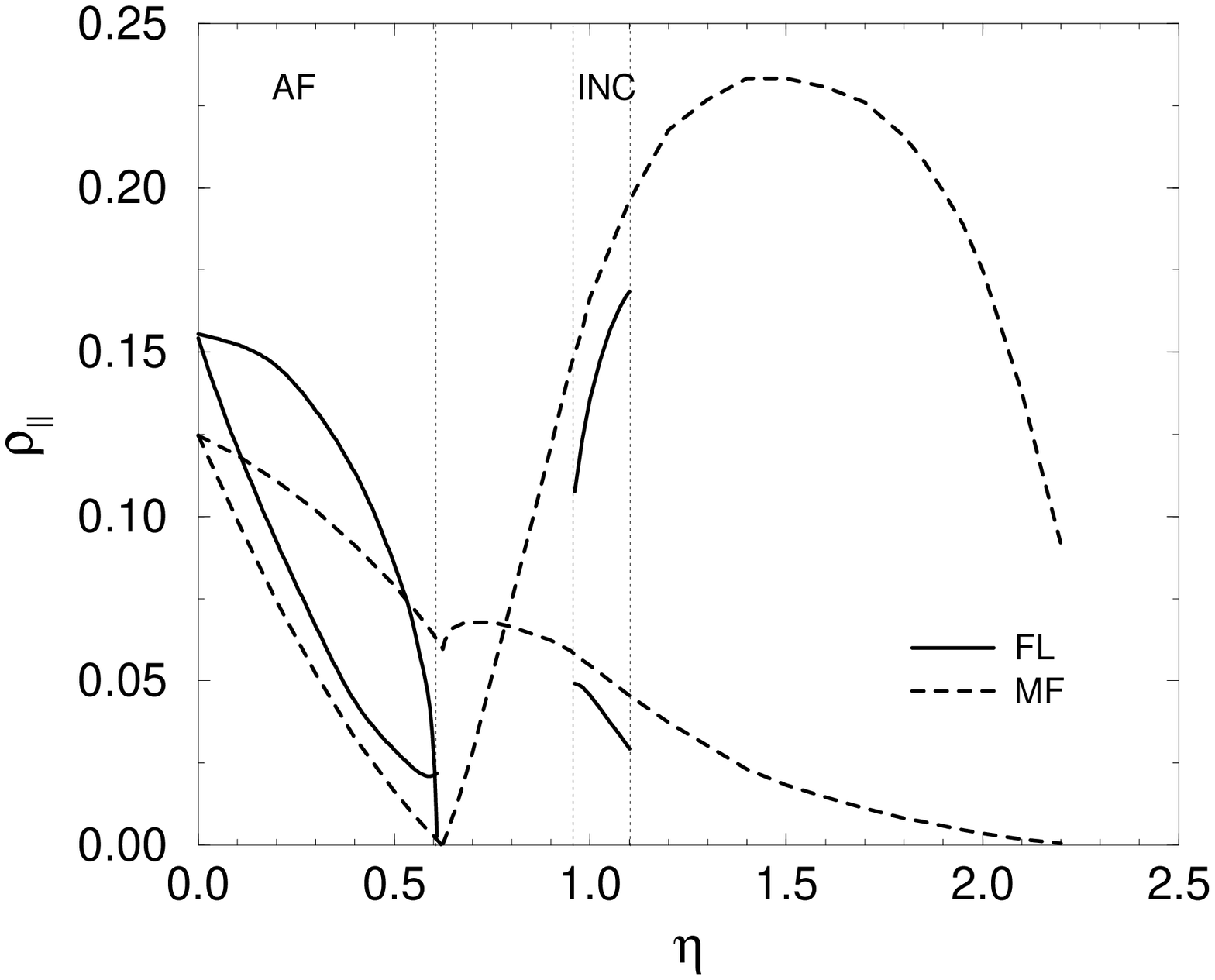}
\caption{}
\end{center}
\end{figure}

\end{document}